\documentclass[twoside,twocolumn,english,prb, showpacs]{revtex4}

\usepackage[latin1]{inputenc}
\usepackage{graphicx}
\usepackage{amssymb}

\makeatletter

\usepackage{babel}
\makeatother
\begin{document}

\title{Equations of motion approach to decoherence and current noise in
ballistic interferometers coupled to a quantum bath}

\author{Florian Marquardt}

\affiliation{Sektion Physik, Center for NanoScience, and Arnold-Sommerfeld-Center
for Theoretical Physics, Ludwig-Maximilians-Universit\"at M\"unchen,
Theresienstr. 37, 80333 Munich, Germany}

\date{19.4.2006}

\email{Florian.Marquardt@physik.lmu.de}

\begin{abstract}
We present a technique for treating many particles moving inside a
ballistic interferometer, under the influence of a quantum-mechanical
environment (phonons, photons, Nyquist noise etc.). Our approach is
based on solving the coupled Heisenberg equations of motion of the
many-particle system and the bath and is inspired by the quantum Langevin
method known for the Caldeira Leggett model. It allows to study decoherence
and the influence of the bath on other properties of the interferometer.
As a first application, we treat a fermionic Mach-Zehnder interferometer.
In particular, we discuss the dephasing rate and present full analytical
expressions for the leading corrections to the current noise, brought
about by the coupling to the quantum bath. 
\end{abstract}

\pacs{73.23.-b, 72.70.+m, 03.65.Yz}

\maketitle

\section{Introduction}

Decoherence, the destruction of quantum-mechanical phase coherence
by a fluctuating environment, plays an important role, ranging from
fundamental questions such as the quantum-classical correspondence
to potential applications (like quantum information). For most of
the recent two decades, the focus of research has been on quantum-dissipative
systems with few degrees of freedom: the most prominent examples are
the single particle (e.g. Caldeira-Leggett model \cite{callegg,Weiss})
or a single two-level system (spin-boson model) and other impurity
models (e.g. the Kondo model in transport through quantum dots). However,
such a description is no longer adequate when it comes to transport
interference effects both in disordered systems (weak localization,
universal conductance fluctuations) or man-made interference devices
(Aharonov-Bohm rings, double quantum dot interferometers, atom chip
interference setups etc.). In those cases, we are dealing with a many-particle
system. As long as this is coupled to classical noise, we can still
use the single particle picture. Both the technical effort and the
physical ideas expand considerably when going over to a full quantum
bath. Up to now, there have been comparatively few treatments of quantum-dissipative
many-particle systems (for examples see \cite{OpenLuttLiquids,aleinerwhichpath,ABring,DDot,FlorianDima}
and references therein).

\begin{figure}
\includegraphics[%
  width=1\columnwidth]{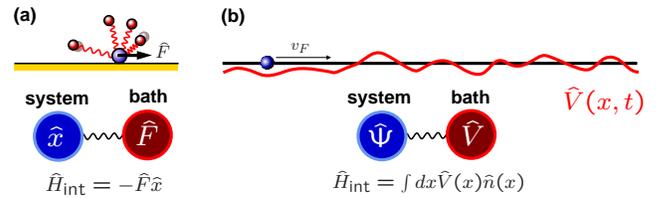}

\caption{\label{cap:Quantum-Langevin-approach}(a) The Caldeira-Leggett model
(single particle and oscillator bath) and (b) a ballistic many-particle
system subject to a quantum noise potential $\hat{V}(x,t)$. }
\end{figure}

In this article, we describe an equations of motion approach for ballistic
interferometers coupled to a quantum bath (Fig. \ref{cap:Quantum-Langevin-approach}).
It is physically transparent, more efficient than generic methods
(like Keldysh diagrams) and straightforwardly keeps important physics,
such as the effects of Pauli blocking in fermionic systems. It may
be applied to describe decoherence (or dephasing; we use the terms
interchangeably) and, in general, to calculate the current noise and
other higher-order correlators of the particle field.

We have already introduced this method in a recent short article\cite{EPL},
and applied it to the electronic Mach-Zehnder interferometer realized
at the Weizmann institute\cite{HeiblumEtAl}, discussing the loss
of visibility in the current interference pattern and (briefly) the
effects on the current noise. The purpose of the present paper is
four-fold: (i) to relate our many-particle method to the Quantum Langevin
equation as it is known for a single particle in the context of the
Caldeira-Leggett model (Sec. \ref{sub:Brief-reminder-of}), (ii) to
provide more details of the method and on how to evaluate the resulting
expressions using perturbation theory (Sec. \ref{sub:Evaluation-of-current}),
(iii) to derive and present full analytical expressions for the leading
correction to the current noise of a MZ setup coupled to a quantum
bath (Sec. \ref{sub:Current-noise-corrections}), and (iv) to add
to our previous brief discussion of the current noise (Sec. \ref{sub:Discussion-of-current}).

\section{Equations of motion approach to decoherence in ballistic interferometers}

\subsection{Brief reminder of the quantum Langevin equation}

\label{sub:Brief-reminder-of}The quantum Langevin equation can be
employed to solve the Caldeira-Leggett model\cite{callegg,Weiss}
of a single particle coupled to a bath of harmonic oscillators. Briefly,
the idea is the following, when formulated on the level of Heisenberg
equations (where it is formally exact). The total quantum force $\hat{F}$
acting on the given particle, due to the bath particles, can be decomposed
into two parts: 

\begin{equation}
\hat{F}(t)=\hat{F}_{(0)}(t)+\int_{-\infty}^{t}D^{R}(t-t')\hat{x}(t')dt'\label{ForceEquation}\end{equation}
The first describes the intrinsic fluctuations, present even in absence
of the coupling. It derives from the solution to the free equations
of motion of the bath oscillators, with fluctuations due to the stochastic
initial conditions. For example, the force might be a linear superposition
of normal oscillator coordinates, $\hat{F}_{(0)}(t)=\sum_{j}g_{j}\hat{Q}_{j(0)}(t)$
with $\hat{Q}_{j(0)}(t)=\hat{Q}_{j(0)}(0)\cos(\Omega_{j}t)+(\hat{P}_{j(0)}(0)/M_{j}\Omega_{j})\sin(\Omega_{j}t)$.
The fluctuations of $\hat{Q}_{j(0)}(0)$ and $\hat{P}_{j(0)}(0)$
includes both thermal and quantum (zero-point) fluctuations. The second
part of the force is due to the response of the bath to the particle's
motion (here: the $\hat{x}$-coordinate, if the coupling is of the
type $\hat{F}\hat{x}$). We will call it the {}``back-action'' term,
and it gives rise to features such as mass renormalization and friction.
As the bath equations of motion are linear (since we are dealing with
a harmonic oscillator bath), the response is linear for arbitrary
coupling strength, and the resulting equation (\ref{ForceEquation})
for the force is valid on the operator level (not only for averages).
In this way, one has {}``integrated out'' the bath by solving for
its motion. Plugging the force $\hat{F}$ into the right-hand-side
(rhs) of the Heisenberg equation of motion for $\hat{x}$ yields the
quantum Langevin equation:

\begin{equation}
m\frac{d^{2}\hat{x}(t)}{dt^{2}}=\hat{F}(t)-U'(\hat{x}(t))\end{equation}
In practice, this equation can only be solved for a harmonic potential
$U(x)$, i.e. for a free particle or a harmonic oscillator. This is
why the range of applications of the quantum Langevin equation is
usually rather restricted. For the example of a harmonic oscillator
(bare frequency $\omega_{0}$), we have, with the help of Eq. (\ref{ForceEquation})
and after going to frequency space: 

\begin{equation}
\hat{x}(\omega)=\frac{\hat{F}_{(0)}(\omega)}{m(\omega_{0}^{2}-\omega^{2})-D^{R}(\omega)}\end{equation}
One can obtain averages of moments of $\hat{x}$ (and $\hat{p}$)
by plugging in the formal solution and employing the correlator of
$\hat{F}_{(0)}(t)$ (using the Wick theorem for higher-order correlators). 

In the case of a many-particle system, it is the density $\hat{n}(x)=\hat{\psi}^{\dagger}(x)\hat{\psi}(x)$
that couples to a scalar noise potential $\hat{V}(x)$. The place
of $\hat{x}$ and $\hat{F}$ in the quantum Langevin equation for
a single particle is thus taken by the particle field $\hat{\psi}$
and $\hat{V}$, respectively.

\subsection{Coupled equations for the many-particle system and the bath}

Let us now turn to the case of many particles (fermions or bosons)
traveling ballistically inside the arm of an interferometer. We will
assume chiral motion and use a linearized dispersion relation, as
this is sufficient to describe decoherence (neglecting acceleration/retardation
effects). We start from Heisenberg's equations of motion for the particles
and the bath. A particle field moving ballistically at constant speed
(see Fig.\ref{cap:Quantum-Langevin-approach} (b)) obeys the following
equation:

\begin{equation}
i(\partial_{t}-v_{F}\partial_{x})\hat{\psi}(x,t)=\int dx'K(x-x')\hat{V}(x',t)\hat{\psi}(x',t)\,,\label{PsiEqMotion}\end{equation}
where $\hat{V}$ evolves in presence of the interaction, see below.
Here $v_{F}$ would be the Fermi velocity in the case of fermions,
or the velocity with which bosons have been injected into the interferometer
(e.g. as a BEC cloud in an atom chip, or the speed of light for photons).
We must consider states within a finite band, thus $K(x-x')=\{\hat{\psi}(x),\hat{\psi}^{\dagger}(x')\}\neq\delta(x-x')$
(written for fermions, analogous for bosons). Nevertheless, for the
purpose of our subsequent leading-order approximation, it turns out
we can replace the right-hand side by $\hat{V}(x,t)\hat{\psi}(x,t)$
(neglecting, e.g., velocity-renormalization in higher orders). The
corresponding formal solution describes the accumulation of a random
{}``quantum phase'':

\begin{eqnarray}
\hat{\psi}(x,t) & = & \hat{T}\exp\left[-i\int_{t_{0}}^{t}dt_{1}\,\hat{V}(x-v_{F}(t-t_{1}),t_{1})\right]\times\nonumber \\
 &  & \hat{\psi}(x-v_{F}(t-t_{0}),t_{0})\,.\label{SolvedPsiEq}\end{eqnarray}
In contrast to the case of classical noise\cite{unserPRL}, the field
$\hat{V}$ contains the response to the particle density, in addition
to the homogeneous solution $\hat{V}_{(0)}$ of the equations of motion
(i.e. the free fluctuations): 

\begin{equation}
\hat{V}(x,t)=\hat{V}_{(0)}(x,t)+\int_{-\infty}^{t}dt'\, D^{R}(x,t,x',t')\hat{n}(x',t')\,.\label{SolvedVEq}\end{equation}
Here $D^{R}$ is the unperturbed retarded bath Green's function, $D^{R}(1,2)\equiv-i\theta(t_{1}-t_{2})\left\langle [\hat{V}(1),\hat{V}(2)]\right\rangle $,
where $\hat{V}$-correlators refer to the free field. This (exact)
step is analogous to the derivation of an operator quantum Langevin
equation, see above. Together with (\ref{SolvedPsiEq}), it correctly
reproduces results from lowest-order diagrammatic perturbation theory.

\begin{figure}
\includegraphics[%
  width=1\columnwidth]{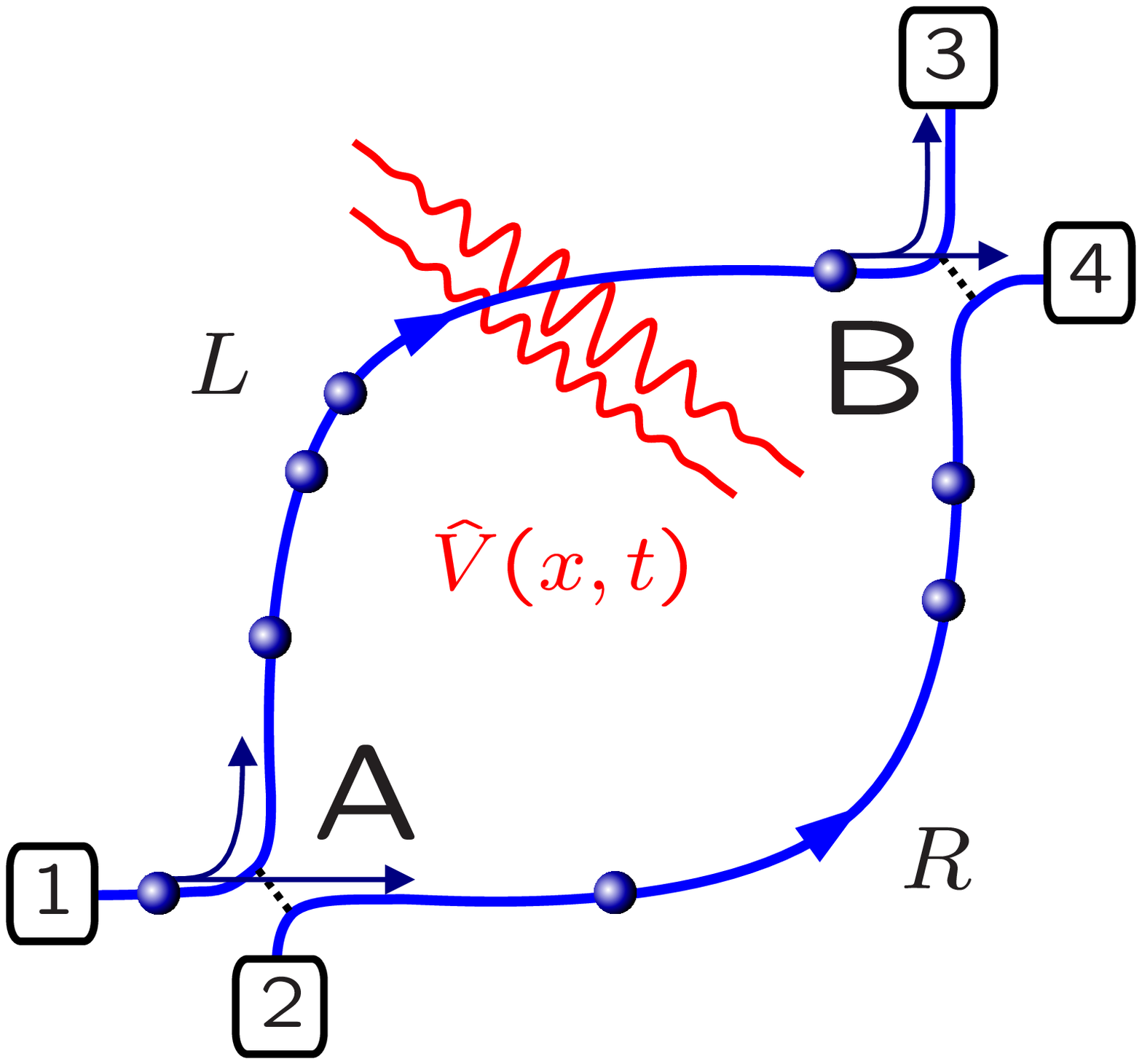}

\caption{\label{cap:Schematic-of-the}Schematic of the Mach-Zehnder setup,
with beam splitters $A,B$, input ports $1,2$, and output ports $3,4$. }
\end{figure}

Below we will apply our approach to the fermionic Mach-Zehnder interferometer,
though the influence of quantum noise on bosonic interferometers (like
in \cite{LevinsonEMField}) represents another interesting future
avenue of research. We note that recently a different kind of quantum
Langevin method has been developed for transport through quantum dots\cite{BingDong}.

\section{Application to the Mach-Zehnder interferometer}

\subsection{Introduction}

In contrast to the usual mesoscopic Aharonov-Bohm ring setups, the
recently realized Mach-Zehnder interferometer for electrons\cite{HeiblumEtAl,NederHeiblum}
offers the possibility to study an ideal two-way interference geometry,
with chiral single-channel transport and in the absence of backscattering.
The loss of visibility with increasing bias voltage or temperature
has been observed, and the idea of using shot noise measurements to
learn more about potential dephasing/decoherence mechanisms has been
introduced. Recent experimental results present a puzzling picture
(e.g. oscillations in the visibility\cite{NederHeiblum}), that has
not been explained so far. Under the assumption that at least part
of the loss in visibility is due to decoherence processes, the observed
decrease in visibility with increasing bias voltage is a good indication
that Pauli blocking effects are important, as this effect is due to
lifting the restrictions of Pauli blocking on the scattering of particles. 

On the theoretical side, the loss of interference contrast in the
current had been studied for the Mach-Zehnder setup\cite{Seelig}
prior to this experiment. More recently the influence of decoherence
on shot noise has been analyzed\cite{unserPRL} (see \cite{levinson}
for related work in quantum point contact), revealing important differences
between phenomenological and microscopic approaches, leading to renewed
investigations on the so-called dephasing terminal model \cite{DephasingTerminal}
and calculations of the full counting statistics in the MZ setup \cite{FCSMZ}.
However, all of these works deal with a classical noise field acting
on the electrons. Thus, experimentally observed features such as the
increase of the dephasing rate with rising bias voltage could not
be studied, as this is a true many-body effect (see below).

\subsection{The model}

We consider a model of spinpolarized fermions, moving chirally and
without backscattering through an interferometer at constant speed
$v_{F}$ (see Fig. \ref{cap:Schematic-of-the}). The two beamsplitters
$A$ and $B$ connect the fermion fields $\hat{\psi}_{\alpha}$ of
the input ($\alpha=1,2$) and output ($\alpha=3,4$) channels to those
of the left and right arm ($\alpha=L,R$), which we take to be of
equal length $l$:

\begin{eqnarray}
\hat{\psi}_{L}(0,t) & = & r_{A}\hat{\psi}_{1}(0,t)+t_{A}\hat{\psi}_{2}(0,t)\label{psiL}\\
\hat{\psi}_{R}(0,t) & = & t_{A}\hat{\psi}_{1}(0,t)+r_{A}\hat{\psi}_{2}(0,t)\label{psiR}\\
\hat{\psi}_{3}(l,t) & = & r_{B}e^{i\phi}\hat{\psi}_{L}(l,t)+t_{B}\hat{\psi}_{R}(l,t)\label{psi3}\\
\hat{\psi}_{4}(l,t) & = & t_{B}e^{i\phi}\hat{\psi}_{L}(l,t)+r_{B}\hat{\psi}_{R}(l,t)\label{psi4}\end{eqnarray}
The transmission (reflection) amplitudes $t_{A/B}\,(r_{A/B})$ fulfill
$t_{j}^{*}r_{j}=-t_{j}r_{j}^{*}$ due to unitarity, and we have included
the Aharonov-Bohm phase difference $\phi$. The input fields $\alpha=1,2$
are described by Fermi distributions $f_{1,2}$, where the chemical
potential difference defines the transport voltage: $eV=\mu_{1}-\mu_{2}$.
We have

\begin{equation}
\left\langle \psi_{\alpha}^{\dagger}(0,0)\psi_{\alpha}(0,t)\right\rangle =\int_{-k_{c}}^{k_{c}}(dk)\, f_{\alpha k}e^{-iv_{F}kt}\label{avgF}\end{equation}
(note $\hbar=1$), with a band-cutoff $k_{c}$. Here and in the following,
we use the notation $(dk)\equiv dk/(2\pi)$. 

The particles are assumed to have no intrinsic interaction, but are
subject to an external free bosonic quantum field $\hat{V}$ (linear
bath) during their passage through the arms $L,R$: $\hat{H}_{{\rm int}}=\sum_{\lambda=L,R}\int dx\,\hat{V}_{\lambda}(x)\hat{n}_{\lambda}(x)$
with $\hat{n}_{\lambda}(x)=\hat{\psi}_{\lambda}^{\dagger}(x)\hat{\psi}_{\lambda}(x)$. 

We focus on the current going into output port $3$, which is related
to the density: $\hat{I}(t)=ev_{F}\hat{n}_{3}(t)$ with $\hat{n}_{3}(t)=\hat{\psi}_{3t}^{\dagger}\hat{\psi}_{3t}$,
where we take fields $\hat{\psi}_{\alpha t}=\hat{\psi}_{\alpha}(l,t)$
at the position of the final beamsplitter B (except where noted otherwise).
In the following we set $e=v_{F}=1$, except where needed for clarity.
We thus have

\begin{equation}
\left\langle \hat{I}\right\rangle =R_{B}\left\langle \hat{\psi}_{L}^{\dagger}\hat{\psi}_{L}\right\rangle +T_{B}\left\langle \hat{\psi}_{R}^{\dagger}\hat{\psi}_{R}\right\rangle +e^{i\phi}t_{B}^{*}r_{B}\left\langle \hat{\psi}_{R}^{\dagger}\hat{\psi}_{L}\right\rangle +{\rm c.c.}.\label{CurrentAverage}\end{equation}
Therefore, the calculation of the average current has been reduced
to a calculation of the elements of a density matrix $\left\langle \hat{\psi}_{\lambda'}^{\dagger}\hat{\psi}_{\lambda}\right\rangle $
describing the coherence properties of the fermions right at the second
beam splitter (after having been subject to the quantum noise field).
We have set $T_{B}=|t_{B}|^{2}$ and $R_{B}=1-T_{B}$.

\subsection{Influence on the interference contrast }

\label{sub:Influence-on-the}

In this section, we will remind the reader of our results for the
influence of the quantum bath on the interference term in the current
$I(\phi)$. These have already been presented in a brief communication\cite{EPL},
but we repeat them here in order to keep the discussion self-contained.
They form the basis of the subsequent sections on the current noise.

In order to obtain the interference term in the current, we expand
the exponential, Eq. (\ref{SolvedPsiEq}), to second order, insert
the formal solution, Eq. (\ref{SolvedVEq}), and perform Wick's averaging
over fermion fields, while implementing a {}``Golden Rule approximation'',
i.e. keeping only terms linear in the time-of-flight $\tau$. 

These steps will be explained in more detail below, in Section \ref{sub:Evaluation-of-current},
for the case of the current noise, so we do not display them here.

Note that accounting for cross-correlations between the fluctuations
in both arms ({}``vertex-corrections'') is straightforward for a
geometry with symmetric coupling to parallel arms at a distance $d$
(assuming $d\ll l$). Then, in the following results (Eqs. (\ref{PhaseShift}),
(\ref{dC0})-(\ref{dC2imag}), and $\Gamma_{\varphi}$), we have to
set $\left\langle \hat{V}\hat{V}\right\rangle =\left\langle \hat{V}_{L}\hat{V}_{L}\right\rangle -\left\langle \hat{V}_{L}\hat{V}_{R}\right\rangle $
and $D^{R}=D_{LL}^{R}-D_{LR}^{R}$. These correlators, of fields being
defined on the one-dimensional interferometer arms, actually have
to be derived from their threedimensional versions, e.g. $\left\langle \hat{V}_{L}(x,t)\hat{V}_{R}(x',t)\right\rangle =\left\langle \hat{V}(x,y+d,z,t)\hat{V}(x',y,z,t')\right\rangle $
if the arms are parallel to the $x$-axis and separated in the $y$-direction.

Without bath, the interference term is given by 

\begin{equation}
\left\langle \hat{\psi}_{R}^{\dagger}\hat{\psi}_{L}\right\rangle _{(0)}=r_{A}t_{A}^{*}\int(dk)\delta f_{k}=r_{A}t_{A}^{*}(eV/2\pi),\label{WObathINTF}\end{equation}
where we define $\delta f_{k}\equiv f_{1k}-f_{2k}$ and $\bar{f}_{k}\equiv(f_{1k}+f_{2k})/2$
(for later). 

The leading correction to the interference term can be expressed in
terms of a phase-shift and a dephasing rate:

\begin{equation}
\delta\left\langle \hat{\psi}_{R}^{\dagger}\hat{\psi}_{L}\right\rangle =r_{A}t_{A}^{*}\int(dk)\delta f_{k}[i\delta\bar{\varphi}(k)-\Gamma_{\varphi}(k)\tau]\label{InterferenceTerm}\end{equation}
Note that the {}``classical'' contributions $\left\langle \hat{\psi}_{\lambda}^{\dagger}\hat{\psi}_{\lambda}\right\rangle $
(with $\lambda=L,R$) are not affected by the noisy environment. Here
the effective average phase shift induced by coupling to the bath
is energy-dependent, and given by:

\begin{equation}
\delta\bar{\varphi}(k)=\tau(R_{A}-T_{A})\int(dq)({\rm Re}D_{q,q}^{R}-D_{0,0}^{R})\delta f_{k-q}\,.\label{PhaseShift}\end{equation}
Essentially, the phase shift is due to the effective coupling between
the electrons, mediated by the bath (containing Hartree and Fock contributions).
For that reason, it depends on the nonequilibrium Fermi distribution
(difference) $\delta f$. The phase shift vanishes for $T_{A}=1/2$,
since then there is complete symmetry between both arms. 

The suppression of the interference term is quantified by the dephasing
rate $\Gamma_{\varphi}(k)$, within the Markoff/Golden Rule approximation
adopted here. In the case of classical Gaussian noise, the suppression
can be evaluated exactly ({}``to all orders'' in the system-bath
interaction). It is equal to $\exp(-\left\langle \varphi^{2}\right\rangle /2)$,
where $\varphi$ is the phase difference between the two arms of the
interferometer, fluctuating due to the action of the noisy potential.
For the case of a \emph{single} particle coupled to a \emph{quantum}
bath, the same suppression factor would be given in general by the
overlap of bath states that have evolved under the influence of the
particle traveling along the left or the right arm\cite{SAI}. Up
to now, we have not been able to find an equally simple interpretation
for the many-particle case.

The total dephasing rate is $\Gamma_{\varphi}(k)=\Gamma_{\varphi}^{L}(k)+\Gamma_{\varphi}^{R}(k)$.
For equal coupling to both arms, this can be written as:

\begin{equation}
\Gamma_{\varphi}(\epsilon)=\int_{0}^{\infty}\frac{d\omega}{v_{F}}{\rm DOS}_{q}(\omega)[2n(\omega)+1-(\bar{f}(\epsilon-\omega)-\bar{f}(\epsilon+\omega))]\label{GammaPhi}\end{equation}
The rate (at energy $\epsilon=\epsilon(k)=v_{F}(k-k_{F})+\epsilon_{F}$)
is an integral over all possible energy transfers $\omega$ from and
to the bath (which have been combined, so $\omega>0$ here). They
are weighted by the bath spectral {}``density of states'' ${\rm DOS}_{q}(\omega)=-{\rm Im}D_{q}^{R}(\omega)/\pi$,
where $q=\omega/v_{F}$ for ballistic motion (in this definition,
${\rm DOS}$ has the dimensions $\omega/q$). 

\begin{figure}
\includegraphics[%
  width=1\columnwidth]{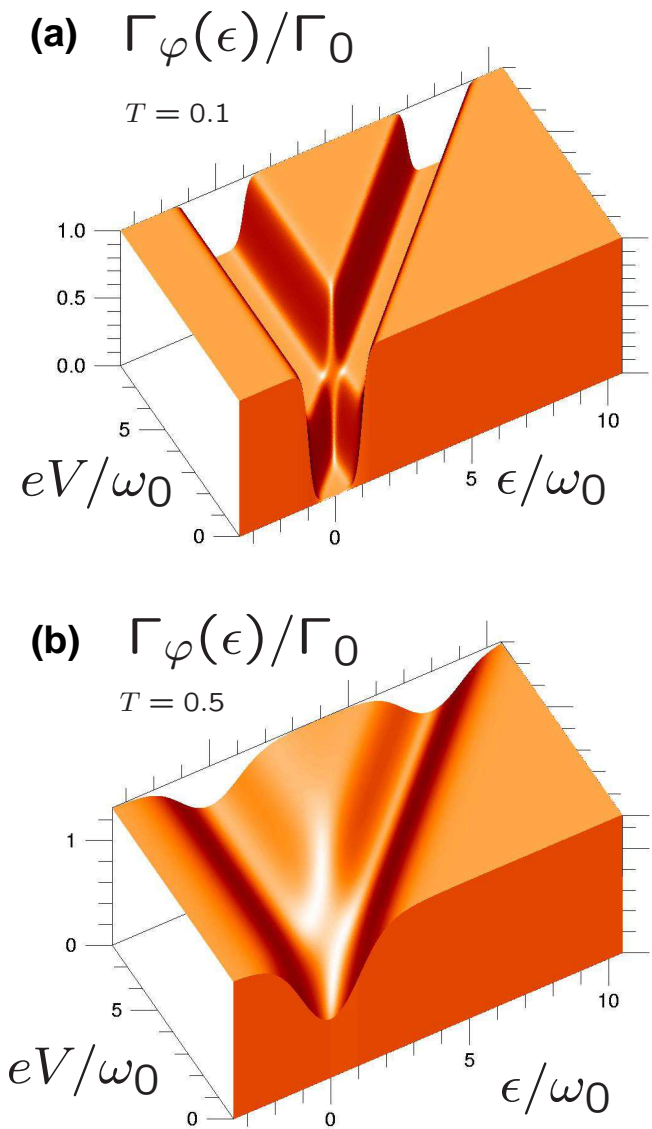}

\caption{\label{cap:Left:-Energy-resolved-dephasing}Energy-resolved dephasing
rate for an optical phonon mode (at $\omega_{0}$), as a function
of transport voltage applied to the Mach-Zehnder, for two different
temperatures: (a) $T=0.1\omega_{0}$, (b) $T=0.5\omega_{0}$. Here
$\Gamma_{0}=\Gamma_{\varphi}(\epsilon\rightarrow\infty,T=0)$.}
\end{figure}

The first term in brackets, $2n(\omega)+1$, describes the strength
of thermal and quantum fluctuations (with $n(\omega)=(e^{\omega/T}-1)^{-1}$
the Bose-Einstein distribution). It stems from the $\hat{V}_{(0)}$
in the quantum phase. By itself, this would give rise to an energy-independent
rate and a visibility independent of bias voltage, in contradiction
to experimental results. In fact, such a procedure (dropping the back-action
terms) would describe a different physical situation: that of a single
particle coupled to a quantum bath (in absence of the Fermi sea). 

Thus, the second term is crucially important. The {}``back-action''
$\propto D^{R}\hat{n}$ introduces the nonequilibrium Fermi functions
($f_{L}=R_{A}f_{1}+T_{A}f_{2}$, $f_{R}=T_{A}f_{1}+R_{A}f_{2}$, and
their average, $\bar{f}=(f_{L}+f_{R})/2=(f_{1}+f_{2})/2$) which capture
the physics of Pauli blocking: Large energy transfers $v_{F}|q|\gg eV,T$
are forbidden for states $k$ within the transport voltage window.
This can be seen in Fig. \ref{cap:Left:-Energy-resolved-dephasing},
which displays the energy-dependence of the dephasing rate, as a function
of voltage and temperature. For the simplest example of an optical
phonon mode (where only an energy transfer $\omega_{0}$ is allowed),
we find two dips in the dephasing rate at large voltages. These occur
around the edges of the non-equilibrium Fermi distribution $\bar{f}$,
i.e. at the edges of the voltage window, and their width is $2\omega_{0}$.
When the voltage is reduced, these two dips merge and the rate goes
down to zero. Thus, when averaging this rate over the voltage window
(in which electrons contribute to the current), the average rate becomes
zero for $V,T\rightarrow0$. As a result, the interference contrast
(visibility) becomes perfect (see also the energy-averaged dephasing
rate depicted in \cite{EPL}). In contrast, at higher temperature,
two effects increase the dephasing rate: First, thermal smearing of
the Fermi distributions reduces the restrictions of Pauli blocking,
and second, thermal fluctuations in the bath lead to processes of
induced emission and absorption.

Note that the strong energy-dependence of the dephasing rate in the
many-fermion case is markedly different from the single-particle situation,
and thus the dependence on the bath spectrum is completely different
as well. In the single-particle case, it is enough to know the variance
$\left\langle \varphi^{2}\right\rangle $ of the fluctuating phase
difference, in order to calculate the loss of visibility. In the many-particle
case, we have to keep track of the full bath spectrum $\left\langle \hat{V}\hat{V}\right\rangle _{q,\omega}$. 

As we have only evaluated the corrections to lowest order, we should
be able to make contact to Fermi's Golden Rule, describing the scattering
of electrons inside the interferometer arms, by emission or absorption
of phonons (bath quanta). Indeed, it turns out that the dephasing
rate is related to Golden Rule scattering rates. However, we emphasize
that it is not given solely by the rate for scattering of particles,
as one might naively assume. Rather, hole-scattering processes provide
an equally important contribution to the dephasing rate, which is
thus the sum of particle- and hole-scattering rates. In our case,
we find:

\begin{equation}
\Gamma_{\varphi}^{L/R}=(\Gamma_{p}^{L/R}+\Gamma_{h}^{L/R})/2,\end{equation}
with $\Gamma_{p}^{L/R}(k)=\int(dq)\,\left\langle \hat{V}\hat{V}\right\rangle _{q,q}(1-f_{L/R,k-q})$
and $\Gamma_{h}^{L/R}(k)=\int(dq)\,\left\langle \hat{V}\hat{V}\right\rangle _{q,q}f_{L/R,k+q}$.
This is because both processes destroy the superposition of many-particle
states that is created when a particle passes through the first beam
splitter, entering the left or the right arm. A more detailed qualitative
discussion may be found in \cite{JvDAndMe}, for the case of weak
localization, and in the next subsection.

For linear transport, i.e. a the limit of infinitesimal bias voltage
$V\rightarrow0$, we have $f_{Lk-q}-f_{Lk+q}\rightarrow-\tanh(\beta(k-q)/2)$
under the integral. Then we recover the result well known in the theory
of weak localization\cite{WLdephasingDiags}, where ballistic motion
in our case ($\omega=v_{F}q$) is replaced by diffusion. 

\begin{figure}
\includegraphics[%
  width=1\columnwidth]{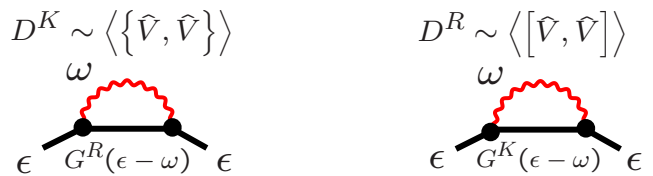}

\caption{\label{KeldyshDiagrams}Contributions to the decoherence rate in
a Keldysh-diagrammatic treatment. Left: Diagram involving both thermal
and quantum fluctuations of the bath, but no Fermi distributions.
This diagram is the same in a single-particle situation. Right: Diagram
corresponding to the {}``back-action'' term discussed in the equations
of motion approach. It involves the fermionic Keldysh Green's function
(that contains the Fermi distribution, $G^{K}(\epsilon-\omega)\propto1-2f(\epsilon-\omega)$)
and the bath's retarded propagator (describing the response).}
\end{figure}

Finally, we note that a treatment using Keldysh diagrams would yield
(in the absence of vertex corrections) a dephasing rate that is equal
to the decay rate of the retarded (or advanced) propagator, and thus
given by the sum of the two diagrams shown in Fig. \ref{KeldyshDiagrams}.
These correspond exactly to the first and the second contribution
discussed above. For the average current, the effort involved in both
calculations (Keldysh or equations of motion) is still about the same
(a few lines). However, for the shot noise corrections discussed below,
we found the equations of motion method much more convenient.

\subsection{Particle- and hole-scattering contributions to the dephasing rate}

In this section, we briefly provide a more qualitative discussion
of the fact that hole-scattering processes lead to an equally important
contribution to the dephasing rate $\Gamma_{\varphi}=(\Gamma_{p}+\Gamma_{h})/2$.
The ratio of $\Gamma_{p}$ and $\Gamma_{h}$ depends on the energy
under consideration, with $\Gamma_{p}$ providing the full dephasing
rate at high energies, and $\Gamma_{h}$ accounting for $\Gamma_{\varphi}$
at low energies (see \cite{EPL}).

\begin{figure}
\includegraphics[%
  width=1\columnwidth]{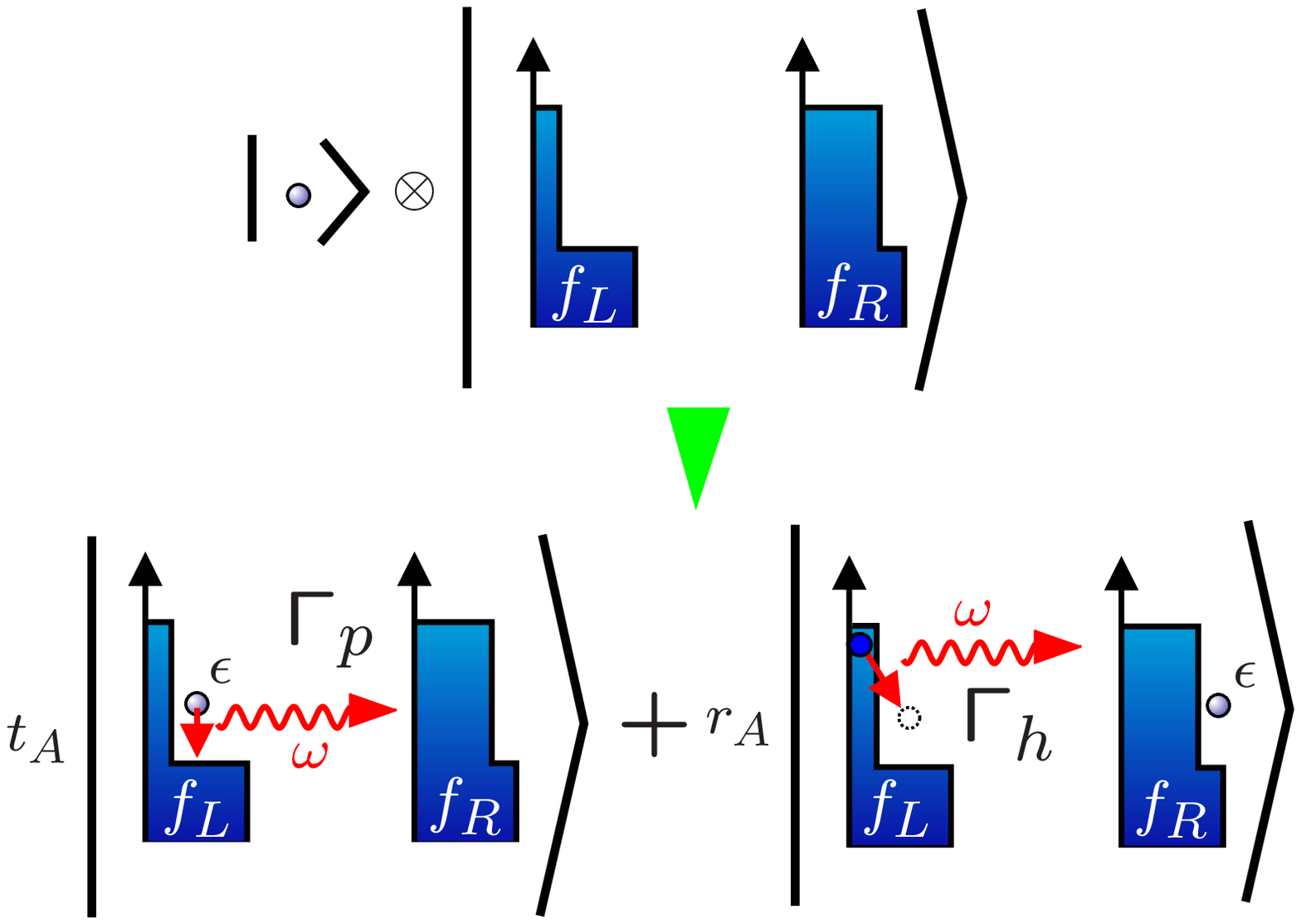}

\caption{\label{cap:Contribution-of-particle-}Contribution of particle- and
hole-scattering processes to the dephasing rate in a many-fermion
interferometer.}
\end{figure}

This is a generic feature for decoherence of fermionic systems. Even
though it is implicit in known diagrammatic results\cite{WLdephasingDiags},
we are not aware of any simple physical discussion (except our own
recent treatment \cite{JvDAndMe} in the case of weak localization).
From the perspective of a single particle, the first beam splitter
creates a superposition of the form $t_{A}\left|R\right\rangle +r_{A}\left|L\right\rangle $,
with the states $R/L$ denoting a wave packet inside the right/left
arm. In the presence of a sea of other fermions inside the interferometer
arms, we should write instead a superposition of many-body states
(see Fig. \ref{cap:Contribution-of-particle-}), schematically:

\begin{equation}
t_{A}\left|\ldots,\underline{0},\ldots;\,\ldots,\underline{1},\ldots\right\rangle +r_{A}\left|\ldots,\underline{1},\ldots;\,\ldots,\underline{0},\ldots\right\rangle \end{equation}
We have indicated the occupations $\left|{\rm left};\,{\rm right}\right\rangle $
of single-particle states in both arms, with a bar denoting the energy
level $\epsilon$ of interest and the remaining particles (in the
nonequilibrium distributions) playing the role of spectators. The
interference term $\left\langle \hat{\psi}_{L}^{\dagger}\hat{\psi}_{R}\right\rangle $
is sensitive to the coherent superposition that requires not only
the presence of a particle in one arm but also the absence of a particle
in the other arm. This is why the many-body superposition can equally
be destroyed by particle- and hole-scattering (leading to states with
$\left|\ldots,\underline{0},\ldots;\,\ldots,\underline{0},\ldots\right\rangle $
or $\left|\ldots,\underline{1},\ldots;\,\ldots,\underline{1},\ldots\right\rangle $,
respectively). We emphasize that the dephasing rate is independent
of the amplitudes $t_{A}$ and $r_{A}$ in this superposition. The
reason is basically that the dephasing rate describes the decay of
the off-diagonal element of the density matrix (in the space of these
two states), and that the amplitudes only enter as a constant prefactor
in that element. Thus, the dephasing rate is simply given by the sum
of particle- and hole-scattering rates, as noted above. The factor
$1/2$ arises because we are not asking about the decay of populations
(which is described by $\Gamma_{p}$ and $\Gamma_{h}$) but essentially
the decay of a wave amplitude. This is the same factor that arises
in the relation $T_{2}=2T_{1}$ known for pure dephasing processes
in the context of Bloch equations.

\section{Current Noise in the Mach-Zehnder setup}

\subsection{Introduction}

As our method yields directly the modified particle fields, it may
be used, in principle, to calculate any higher-order correlator of
those fields. Of particular experimental interest is the current noise
in the output port of the interferometer. This has been (and is) currently
being studied in the Weizmann MZ setup\cite{HeiblumEtAl,NederHeiblum}.

\subsection{General properties}

The zero-frequency current noise power is defined as

\begin{equation}
S\equiv\int_{-\infty}^{+\infty}dt\,\left\langle \left\langle \hat{I}(t)\hat{I}(0)\right\rangle \right\rangle \,,\label{SDef}\end{equation}
where the double bracket denotes the irreducible part: $\left\langle \left\langle \hat{I}(t)\hat{I}(0)\right\rangle \right\rangle =\left\langle \hat{I}(t)\hat{I}(0)\right\rangle -\left\langle \hat{I}\right\rangle ^{2}$.
For the MZ setup considered here, the current noise only has contributions
up to the second harmonic in the external flux:

\begin{equation}
S=S_{0}+S_{1}\cos(\phi-\delta\phi_{1})+S_{2}\cos(2(\phi-\delta\phi_{2})).\label{Sharmonics}\end{equation}
The dependence on $\phi$ and $T_{B},R_{B}$ can be made explicit, 

\begin{eqnarray}
 &  & S=R_{B}T_{B}C_{0}+R_{B}^{2}C_{0R}+T_{B}^{2}C_{0T}+\label{S}\\
 &  & 2\textrm{Re}\left[e^{i\phi}(t_{B}^{*}r_{B})(R_{B}C_{1R}+T_{B}C_{1T})-e^{2i\phi}T_{B}R_{B}C_{2}\right]\,.\nonumber \end{eqnarray}
with the coefficients following directly from inserting Eq. (\ref{psi3})
into (\ref{SDef}), see below. Here $S_{0},S_{1},S_{2},\delta\phi_{1}$
and $\delta\phi_{2}$ can be obtained by comparing Eqs. (\ref{S})
and (\ref{Sharmonics}).

The coefficients $C_{0},C_{0R},\ldots$ are expressed in terms of
four-point Green's functions, similar to the expression for the average
current. These, in turn, contain the full dependence on interactions,
as well as on voltage, temperature, and $T_{A}$. We list them for
reference, setting $L_{t}\equiv\hat{\psi}_{L}(t)$ and $R_{t}\equiv\hat{\psi}_{R}(t)$
for brevity. We will also set $v_{F}\equiv1$, as before.

\newcommand{\Lt}{L_{t}}

\newcommand{\Ltd}{L_{t}^{\dagger}}

\newcommand{\Rt}{R_{t}}

\newcommand{\Rtd}{R_{t}^{\dagger}}

\newcommand{\Lo}{L_{0}}

\newcommand{\Lod}{L_{0}^{\dagger}}

\newcommand{\Ro}{R_{0}}

\newcommand{\Rod}{R_{0}^{\dagger}}

\begin{eqnarray}
C_{0R} & = & \int dt\,\left\langle \left\langle \Ltd\Lt\Lod\Lo\right\rangle \right\rangle \label{C0R}\\
C_{0T} & = & \int dt\,\left\langle \left\langle \Rtd\Rt\Rod\Ro\right\rangle \right\rangle \\
C_{0} & = & \int dt\,\left\langle \left\langle \Ltd\Lt\Rod\Ro+\Rtd\Rt\Lod\Lo\right\rangle \right\rangle +\\
 &  & \int dt\,\left\langle \left\langle \Ltd\Rt\Rod\Lo\right\rangle \right\rangle +\left\langle \left\langle \Rtd\Lt\Lod\Ro\right\rangle \right\rangle \label{C0}\\
C_{1R} & = & \int dt\,\left\langle \left\langle \Rtd\Lt\Lod\Lo\right\rangle \right\rangle +\left\langle \left\langle \Ltd\Lt\Rod\Lo\right\rangle \right\rangle \label{C1R}\\
C_{1T} & = & \int dt\,\left\langle \left\langle \Rtd\Rt\Rod\Lo\right\rangle \right\rangle +\left\langle \left\langle \Rtd\Lt\Rod\Ro\right\rangle \right\rangle \label{C1T}\\
C_{2} & = & \int dt\,\left\langle \left\langle \Rtd\Lt\Rod\Lo\right\rangle \right\rangle \label{C2}\end{eqnarray}
$C_{0(R/T)}$ are real-valued, the other coefficients may become complex. 

In the absence of a quantum bath, these coefficients have the following
values:

\begin{eqnarray}
C_{0R/T}^{(0)} & = & \int(dk)\left[\bar{f}_{k}(1-\bar{f}_{k})-\frac{1}{4}(R_{A}-T_{A})^{2}\delta f_{k}^{2}\right]\\
C_{0}^{(0)} & = & \int(dk)\left\{ f_{Lk}(1-f_{Rk})+f_{Rk}(1-f_{Lk}))\right\} \\
 &  & -2R_{A}T_{A}\int(dk)\,\delta f_{k}^{2}\label{C00}\\
C_{1R/T}^{(0)} & = & \pm r_{A}t_{A}^{*}(T_{A}-R_{A})\int(dk)\,\delta f_{k}^{2}\,\\
C_{2}^{(0)} & = & R_{A}T_{A}\int(dk)\,\delta f_{k}^{2}\,.\end{eqnarray}
Those expressions yield the result given by the well-known scattering
theory of shot noise of non-interacting fermions \cite{PartitionNoiseOriginal,BuettikerOriginal,JongBeenakker,BlanterBuettiker}:

\begin{equation}
S_{(0)}=\int(dk)(f_{2k}+\delta f_{k}\mathcal{T})(1-(f_{2k}+\delta f_{k}\mathcal{T}))\,,\label{SNfree}\end{equation}
where $\mathcal{T}(\phi)=T_{A}T_{B}+R_{A}R_{B}+2t_{A}^{*}r_{A}t_{B}^{*}r_{B}\cos(\phi)$
is the transmission probability from $1$ to $3$. 

For our model, the full shot noise power $S$ may be shown to be invariant
under each of the following transformations, if the bath couples equally
to both arms of the interferometer: (i) $t_{A}\leftrightarrow r_{A},\,\phi\mapsto-\phi$
(ii) $V\mapsto-V,\,\phi\mapsto-\phi$ (iii) $t_{B}\leftrightarrow r_{B}$.
As a consequence, $C_{1T}=-C_{1R}$. Note that the free result (\ref{SNfree})
is invariant under $\phi\mapsto-\phi$ and $V\mapsto-V$ separately,
but these symmetries may be broken by a bath-induced phase-shift,
to be discussed below.

\subsection{Evaluation of current noise to leading order in the interaction}

\label{sub:Evaluation-of-current}In order to evaluate the correlators
(\ref{C0R})-(\ref{C2}) to leading order in the interaction, we expand
the general solution of the equations of motion for the electron operators.
Let $L_{t}^{(0)}$ denote the unperturbed electron field, and $g$
a formal expansion parameter (to be set to $1$ in the end). Then
we have, for the electron field at the end of the left arm, just before
the final beamsplitter:

\begin{eqnarray}
L_{t} & = & \left[1-ig\int_{0}^{\tau}dt_{1}\tilde{V}_{L}(t_{1},t)-\right.\label{Lexpr}\\
 &  & g^{2}\int_{0}^{\tau}dt_{1}\int_{0}^{t_{1}}dt_{2}\,\tilde{V}_{L}(t_{1},t)\tilde{V}_{L}(t_{2},t)-\nonumber \\
 &  & ig^{2}\int_{0}^{\tau}dt_{1}\int_{-\infty}^{+\infty}dt_{2}\int_{-\infty}^{+\infty}dx_{2}\,\times\nonumber \\
 &  & \left.\tilde{D}_{LL}^{R}(t_{1},t;x_{2},t_{2})\, L_{\tau+t_{2}-x_{2}/v_{F}}^{(0)\dagger}L_{\tau+t_{2}-x_{2}/v_{F}}^{(0)}\right]L_{t}^{(0)}\nonumber \end{eqnarray}
 We have expressed the arguments of the potentials $V$ and response
kernel $D^{R}$ in terms of the time $t_{1}=0\ldots\tau$ elapsed
since entry into the left interferometer arm, with the electron moving
from $x=0$ to $x=l=v_{F}\tau$ during the corresponding time-interval
$[t-\tau,t]$. We have set

\begin{eqnarray}
\tilde{V}_{L}(t_{1},t) & \equiv & \hat{V}_{L}(v_{F}t_{1},t-\tau+t_{1})\\
\tilde{D}_{LL}^{R}(t_{1},t;x_{2},t_{2}) & \equiv & D^{R}(v_{F}t_{1}-x_{2},t-\tau+t_{1}-t_{2})\,,\end{eqnarray}
assuming a stationary environment that is translationally invariant.
The expressions for $R_{t}$ are completely analogous. In writing
down Eq. (\ref{Lexpr}), we have omitted the cross-term $D_{LR}^{R}$,
assuming that the wavelength of relevant fluctuations is considerably
shorter than the distance between the arms of the interferometer (such
a term can be added easily, see the remark above, in Section \ref{sub:Influence-on-the}).
This also implies $\left\langle \hat{V}_{L}\hat{V}_{R}\right\rangle =0$.
In terms of the bath spectra, we have (both for $L$ and $R$):

\begin{eqnarray}
 &  & \left\langle \tilde{V}_{L}(t'_{1},t')\tilde{V}_{L}(t_{1},t)\right\rangle =\nonumber \\
 &  & \int(dq)\int(d\omega)\, e^{i[(v_{F}q-\omega)(t'_{1}-t_{1})-\omega(t'-t)]}\left\langle \hat{V}\hat{V}\right\rangle _{q\omega}\\
 &  & \tilde{D}_{LL}^{R}(t_{1},t;x_{2},t_{2})=\nonumber \\
 &  & \int(dq)\int(d\omega)e^{i[q(v_{F}t_{1}-x_{2})-\omega(t_{1}-t_{2}+t-\tau)]}D_{q\omega}^{R}\end{eqnarray}
We now evaluate the leading order ($g^{2}$) correction to the noise
power (\ref{S}), by inserting the expressions for $L_{t}$ and $R_{t}$
into the coefficients $C_{0},C_{1T},C_{1R},$ and $C_{2}$ (Eqs. (\ref{C0})-(\ref{C2})).
Bare electron operators are contracted using Wick's theorem, and the
resulting averages can be performed by expressing $L_{t}^{(0)},\, R_{t}^{(0)}$
via $\hat{\psi}_{1,2}$ (Eqs. (\ref{psiL}),(\ref{psiR})) and employing
Eq. (\ref{avgF}). After inserting the Fourier representations $\left\langle \hat{V}\hat{V}\right\rangle _{q\omega}$
and $D_{q\omega}^{R}$, all temporal and spatial integrations have
to be carried out. In doing so, we will use a Golden Rule (Markoff)
approximation, i.e. we keep only the leading order in $\tau$,

\begin{eqnarray}
\int_{0}^{\tau}dt_{1}\int_{0}^{t_{1}}dt_{2}e^{i\lambda(t_{1}-t_{2})} & \approx & \tau\frac{i}{\lambda+i0}\,\end{eqnarray}
(and so on), assuming the correlation time of the environment to be
much shorter than the time-of-flight $\tau$. Although it is in principle
straightforward to go beyond this approximation (evaluating all these
integrals exactly), the result gets very unwieldy, and other effects
(such as the curvature of the interferometer paths) should be taken
into account as well on that refined level of description. Thus, we
are neglecting the fact that energy- and momentum-conservation will
only be fulfilled up to a Heisenberg uncertainty $\tau^{-1}$ and
$l^{-1}$, respectively. Within this approximation, we have been allowed
to extend the $x_{2}$-integral in Eq. (\ref{Lexpr}) over all of
space, even though the interaction is assumed to be restricted to
the interferometer arm (it will be restricted automatically by the
short range of $D^{R}$ and the fact that $t_{1}\in[0,\tau]$).

\subsection{Current noise corrections due to the quantum bath}

\label{sub:Current-noise-corrections}After a straightforward but
lengthy calculation, we arrive at the leading-order corrections to
the coefficients $C_{0},C_{1R},C_{2}$ in the noise power $S$. Here
we list the explicit analytical results for the shot noise correction
(cf. Eq. \ref{S}), valid for arbitrary bath spectra (note $\delta C_{0R/T}=0$
and $\delta C_{1T}=-\delta C_{1R}$):

\begin{eqnarray}
 &  & \frac{\delta C_{0}}{4\tau R_{A}T_{A}}=-\int(dk)(dq)\,{\rm Im}D_{q,q}^{R}\times\nonumber \\
 &  & \left[\delta f_{k}\delta f_{k+q}(\bar{f}_{k+q}-\bar{f}_{k})+\right.\nonumber \\
 &  & \left.(f_{1k}^{2}+f_{2k}^{2})\bar{f}_{k+q}-(f_{1k+q}^{2}+f_{2k+q}^{2})\bar{f}_{k}\right]+\nonumber \\
 &  & \int(dk)(dq)\left\langle \hat{V}\hat{V}\right\rangle _{q,q}\times\nonumber \\
 &  & [(f_{1k+q}-f_{1k})(1-f_{1k})+(f_{2k+q}-f_{2k})(1-f_{2k})]\nonumber \\
 &  & +(eV/2\pi)^{2}\left\langle \hat{V}\hat{V}\right\rangle _{0,0}\,,\label{dC0}\end{eqnarray}

\begin{eqnarray}
 &  & {\rm Re}\frac{\delta C_{1R}}{\tau r_{A}t_{A}^{*}(R_{A}-T_{A})}=\int(dk)(dq){\rm Im}D_{q,q}^{R}\times\nonumber \\
 &  & [\delta f_{k}\delta f_{k+q}(\bar{f}_{k+q}+3\bar{f}_{k}-2)+\delta f_{k+q}^{2}\bar{f}_{k}-\delta f_{k}^{2}\bar{f}_{k+q}]+\nonumber \\
 &  & \left[\int(dq)\left\langle \hat{V}\hat{V}\right\rangle _{q,q}\right]\left[\int(dk)\delta f_{k}^{2}\right]\,,\label{dC1Rreal}\end{eqnarray}

\begin{eqnarray}
 &  & {\rm Im}\frac{\delta C_{1R}}{\tau r_{A}t_{A}^{*}}=\int(dk)(dq){\rm Re}D_{q,q}^{R}\times\nonumber \\
 &  & \left[-\delta f_{k}\delta f_{k+q}(\delta f_{k+q}+2\delta f_{k})(T_{A}-R_{A})^{2}/2+\right.\nonumber \\
 &  & 2R_{A}T_{A}\delta f_{k}\delta f_{k+q}^{2}+\delta f_{k+q}\bar{f}_{k}(3-2\bar{f}_{k})\nonumber \\
 &  & \left.-\delta f_{k}\bar{f}_{k+q}+2\bar{f}_{k}\bar{f}_{k+q}(\delta f_{k}-\delta f_{k+q})\right]\nonumber \\
 &  & +D_{0,0}^{R}(eV/2\pi)\int(dk)\times\nonumber \\
 &  & \left[\delta f_{k}^{2}(\frac{3}{2}(T_{A}^{2}+R_{A}^{2})-5R_{A}T_{A})-2\bar{f}_{k}(1-\bar{f}_{k})\right]\,,\label{dC1Rimag}\end{eqnarray}

\begin{eqnarray}
 &  & \frac{{\rm Re}\delta C_{2}}{2\tau R_{A}T_{A}}=(eV/2\pi)^{2}\left\langle \hat{V}\hat{V}\right\rangle _{0,0}-\nonumber \\
 &  & 2\int(dk)(dq)\,{\rm Im}D_{q,q}^{R}\bar{f}_{k}\delta f_{k+q}(\delta f_{k}+\delta f_{k+q})-\nonumber \\
 &  & \int(dk)(dq)\left\langle \hat{V}\hat{V}\right\rangle _{q,q}\delta f_{k}(\delta f_{k}+\delta f_{k+q})\label{dC2real}\end{eqnarray}

\begin{eqnarray}
 &  & {\rm Im}\delta C_{2}=-4R_{A}T_{A}\int(dk)\delta f_{k}^{2}\delta\bar{\varphi}_{k}\,.\label{dC2imag}\end{eqnarray}
To obtain physical insights, it is best to translate the coefficients
that have been obtained above into corrections to the different harmonics
$S_{0},S_{1},S_{2}$ of the noise pattern $S(\phi)$ and the phase
shifts $\delta\phi_{1}$ and $\delta\phi_{2}$ (compare Eqs. (\ref{Sharmonics})
and (\ref{S})). Then we find, for the lowest order corrections:

\begin{eqnarray}
\delta S_{0} & = & R_{B}T_{B}\delta C_{0}\label{dS0}\\
\frac{\delta S_{1}}{S_{1}} & = & \frac{i{\rm Im}\delta C_{1R}}{C_{1R}^{(0)}}\\
\frac{\delta S_{2}}{S_{2}} & = & \frac{{\rm Re}\delta C_{2}}{C_{2}^{(0)}}\\
\delta\phi_{1} & = & \frac{i{\rm Re}\delta C_{1R}}{C_{1R}^{(0)}}\\
\delta\phi_{2} & = & -\frac{{\rm Im}\delta C_{2}}{C_{2}^{(0)}}\label{dPhi2}\end{eqnarray}
(Note, when comparing with Eqs. (\ref{dC1Rreal}) and (\ref{dC1Rimag}),
that we took into account $r_{A}t_{A}^{*}$, and thus also $C_{1R}^{(0)}$,
being purely imaginary)

\subsection{Discussion of current noise in the Mach-Zehnder coupled to a quantum
bath}

\label{sub:Discussion-of-current}

\begin{figure}
\includegraphics[%
  width=1\columnwidth]{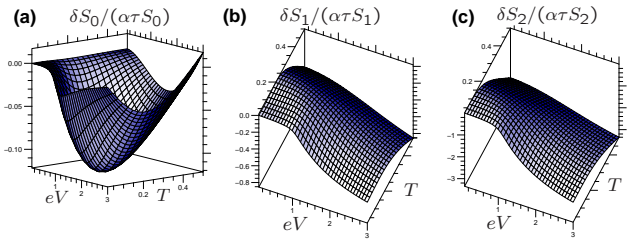}

\caption{\label{cap:(a)-Correction-to}(a) Correction to the flux-averaged
current noise power $S_{0}$ for a damped optical phonon mode (strength
$\alpha$, time-of-flight $\tau$), normalized with respect to unperturbed
value. (b) and (c): The corrections to the first and second harmonics
$S_{1}$ and $S_{2}$ are negative, revealing the loss of interference
contrast in $S(\phi)$. Energies are plotted in units of the phonon
mode ($\omega_{0}=1$), and the MZ setup has been chosen asymmetric
($T_{A}=0.3,T_{B}=0.4$).}
\end{figure}

\begin{figure}
\includegraphics[%
  width=1\columnwidth]{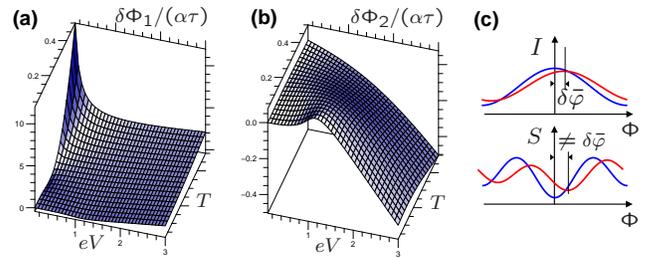}

\caption{\label{cap:Phase-corrections}Corrections to the phase shifts $\delta\phi_{1}$
(a) and $\delta\phi_{2}$ (b) of the first and second harmonic in
the shot noise pattern $S(\phi)$, in an asymmetric MZ setup (parameters
as in Fig. \ref{cap:(a)-Correction-to}). These phase shifts will,
in general, be different from the effective phase shift $\delta\bar{\varphi}$
in the current pattern, leading to the situation schematically depicted
in (c).}
\end{figure}

The results of evaluating Eqs. (\ref{dS0})-(\ref{dPhi2}) are shown
in Figs. (\ref{cap:(a)-Correction-to}) and (\ref{cap:Phase-corrections})
for the illustrative example of a damped optical phonon mode, $D_{q,\omega}^{R}=\alpha[(\omega-\omega_{0}+i\eta)^{-1}-(\omega+\omega_{0}+i\eta)^{-1}]$,
with $\eta/\omega_{0}=0.1$.

As expected, the $\phi$-dependence of the shot noise (\ref{S}) is
suppressed, i.e. not only the visibility (interference contrast) of
the current pattern $I(\phi)$ but also that of the shot noise pattern
$S(\phi)$ is reduced by the bath: see Fig. \ref{cap:(a)-Correction-to}
(b) and (c). We emphasize that this reduction becomes noticeable only
once the voltage or the temperature become comparable to the frequency
of the phonon mode. Only then the particle can lose its coherence
by leaving a trace in the bath (that acts as a kind of {}``which-way
detector''). This is the same behaviour found for the visibility
of the current, and it is satisfying that this simple qualitative
physical idea also holds for decoherence in shot noise. Note, however,
that we have not found a way to express the comparatively complicated
formulas for $\delta S_{1}$ and $\delta S_{2}$ in terms of the simple
dephasing rate which we derived above, Eq. \ref{GammaPhi}. It is
interesting to note that the decrease of the second harmonic $S_{2}$
proceeds faster than that of the first harmonic, $S_{1}$. This is
qualitatively consistent with the observations made by Chung et al.
for a MZ setup using the phenomenological dephasing terminal model\cite{chung}.

There is no Nyquist noise correction, as seen in Fig. \ref{cap:(a)-Correction-to}
(a), at $V=0$. This can be understood easily, since the (unperturbed)
Nyquist noise $S_{(0)}(V=0)$ does not depend on $\phi$ and thus
should not be sensitive to a noisy environment that changes the phase
$\phi$. 

The limit of classical noise (treated to all orders in Refs. \cite{unserPRL})
is recovered by setting $D^{R}=0$ and using the symmetrized correlator
$\left\langle V_{{\rm cl}}V_{{\rm cl}}\right\rangle =\left\langle \left\{ \hat{V},\hat{V}\right\} \right\rangle /2$
everywhere in the shot noise correction derived here, with the exception
of Eq. (\ref{dC0}), which has to be replaced by:

\begin{eqnarray}
 &  & \frac{\delta C_{0}^{{\rm cl}}}{\tau}=2\int(dk)(dq)\left\langle V_{{\rm cl}}V_{{\rm cl}}\right\rangle _{q,q}\times\nonumber \\
 &  & [(f_{Lk+q}-f_{Lk})(1-f_{Rk})+(f_{Rk+q}-f_{Rk})(1-f_{Lk})]+\nonumber \\
 &  & 4R_{A}T_{A}\left[\int(dk)\,\delta f_{k}\right]^{2}\left\langle V_{{\rm cl}}V_{{\rm cl}}\right\rangle _{0,0}\,.\label{dC0classical}\end{eqnarray}
This contribution contains a finite $\phi$-independent Nyquist noise
correction (cf. \cite{unserPRL}), in contrast to our result for the
quantum bath. This may be understood as being due to heating of the
MZ electrons by a bath which is nominally at infinite temperature
(according to the fluctuation-dissipation theorem FDT, applied to
the case $D^{R}=0$).

We emphasize that it is impossible to recover the full quantum noise
result by inserting some suitably modified classical noise correlator
$\left\langle V_{{\rm cl}}V_{{\rm cl}}\right\rangle $. This is in
contrast to the dephasing rate, where such a procedure (with $\left\langle V_{{\rm cl}}V_{{\rm cl}}\right\rangle $
containing Fermi functions for Pauli blocking, see Refs. \cite{JvDAndMe}
or also\cite{CohenImry}) can be made to work. In particular, having
only classical noise cannot yield the important phase shift terms.
In contrast, the conductance fluctuations are correctly captured even
by the classical approach.

At large $V$ (larger than the bath spectrum cutoff), there is a contribution
$\propto V^{2}$ in $\delta S_{0}$ and $\delta S_{2}$, due to time-dependent
conductance fluctuations ($\delta I\propto\delta G(\phi(t))\cdot V$),
corresponding to the leading order of {}``$S_{{\rm cl}}$'' in Refs.
\cite{unserPRL} (see $\left\langle \hat{V}\hat{V}\right\rangle _{0,0}$
terms in Eqs. (\ref{dC0}) and (\ref{dC2real})). 

As mentioned in \cite{EPL} the main surprising feature connected
to the shot noise correction is the behaviour of the phase-shifts
$\delta\phi_{1,2}$. Naively, one might expect the effective phase
shift to be one and the same for all quantities depending on the Aharonov-Bohm
phase, whether it be the current $I(\phi)$ or the shot noise $S(\phi)$.
However, the phase-shift $\delta\phi_{2}$ in the $e^{2i\phi}$ term
is twice as large as expected from the phase-shift in $I(\phi)$ and,
moreover, the $e^{i\phi}$ phase-shift $\delta\phi_{1}$ does not
vanish even if $T_{A}=1/2$ (but $T_{B}\neq1/2$). As a consequence,
and in contrast to the current $I(\phi)$, even for completely symmetric
interferometer arms (same density, same Fermi distributions, same
coupling to the bath), there remains a $\phi\leftrightarrow-\phi$-asymmetry
in $\delta S$. The explanation\cite{EPL} rests on the fact that
the phase shift is sensitive to the density difference between the
arms (as discussed above). As a consequence, density fluctuations
in both arms also lead to fluctuations of this phase shift. While
the average current only feels the average phase shift $\delta\bar{\varphi}$,
the current noise is affected by those fluctuations. The extra terms
in $\delta S$, which are responsible for the deviation from the behaviour
of the average current, come about because the fluctuations of the
phase shift are correlated with the output current, $\left\langle (\delta\hat{\varphi}(t)-\delta\bar{\varphi})(\hat{I}(0)-\bar{I})\right\rangle \neq0$.
This is a straightforward consequence of the fact that the output
current $\hat{I}$ itself is correlated with the currents/densities
traveling inside the interferometer arms. This also explains the fact
that $T_{A}=1/2$ is not enough to obtain a $\phi$-symmetric shot
noise (since the correlator $\left\langle \delta\hat{\varphi}\hat{I}\right\rangle $
depends on $T_{B}$ as well). 

We emphasize that a fluctuating effective phase shift depending on
the density fluctuations inside the arms will quite likely be present
in any model of interacting fermions moving inside an interferometer
(either with intrinsic interactions, i.e. as a Luttinger liquid\cite{LawFeldmanGefen},
or with interactions mediated by a bath, like in the present work).
Thus, the consequences (different phase shifts in current and current
noise, and different phases of the two harmonics in the current noise)
will hold more generally than our specific model. It is thus important
to carry out experiments that test for those phase shifts in asymmetric
interacting interferometers.

\subsection{Conclusions}

We have presented an equations of motion (quantum Langevin) approach
to ballistic interferometers containing many particles coupled to
a quantum bath. It takes into account the simplifications provided
by the chiral motion at approximately constant velocity, and is thus
more efficient than more general approaches. In particular, it is
able to keep, in a straightforward and physically transparent manner,
many-body effects, such as Pauli blocking (described as a consequence
of the backaction of the bath onto the system) or the influence of
hole-scattering processes in the case of fermions. We have applied
this method to the fermionic Mach-Zehnder interferometer, presenting
full analytical results for the influence of the quantum bath on the
current noise. As we have discussed, the main effects are a reduction
of the interference contrast in the shot noise pattern $S(\phi)$
and a peculiar behaviour of the effective phase shifts in the two
harmonics of $S(\phi)$, for asymmetric setups. 

We are anticipating future applications such as the treatment of higher-order
effects of the bath or decoherence in bosonic (atom-chip) interferometers. 

\begin{acknowledgments}
I thank B. Abel, I. Neder, M. Heiblum, U. Gavish, Yu. Gefen, Y. Levinson,
Y. Imry, M. B\"uttiker, S. M. Girvin, A. A. Clerk, C. Bruder, J.
v. Delft, T. Novotn\'{y} and V. Fal'ko for illuminating discussions,
and the DFG and the BMBF for financial support.
\end{acknowledgments}

\end{document}